\documentclass[aps,prl,twocolumn,floatfix]{revtex4}
\usepackage{graphicx}
\usepackage{psfrag}

\begin{document}

\title{Exact coherent states of a harmonically confined Tonks-Girardeau gas}

\author{A.~Minguzzi and D.M.~Gangardt}

\affiliation{\mbox{Laboratoire de Physique 
Th\'eorique et Mod\`eles Statistiques,\\
Universit\'e Paris-Sud, B\^at. 100, F-91405 Orsay, France}}

\date{\today}

\begin{abstract}
Using a scaling transformation we exactly determine the dynamics 
of an harmonically confined Tonks-Girardeau gas under arbitrary 
time variations of the trap frequency. We show  how during 
a one-dimensional expansion a ``dynamical fermionization'' occurs 
as the momentum distribution rapidly approaches an ideal Fermi gas 
distribution, and that under a sudden change of the trap frequency 
the gas undergoes undamped breathing oscillations displaying alternating 
bosonic and fermionic character in momentum space.
The absence of damping in the oscillations is a peculiarity of the truly
Tonks regime.

\end{abstract}

\maketitle

One of the most challenging and interesting problems in quantum
dynamics involves  understanding  a 
temporal behavior of strongly
correlated many-body systems beyond the linear-response regime or the 
adiabatic approximation. Aside from fundamental interest,
this issue is of primary importance for  current experiments 
with ultracold atomic gases and their potential  applications to
quantum information, where control and
manipulation of entangled states of  many-particle quantum systems is required. 

The one-dimensional (1D) gas of impenetrable bosons (Tonks-Girardeau
gas) corresponds to the limit of infinitely strong repulsive
interactions in the Lieb-Liniger (LL)  model of 1D bosons interacting
through a contact pair potential \cite{LiebLiniger1963}. It was shown
in Ref. \cite{Petrov2000} that the Tonks-Girardeau limit is applicable for     
describing the low-density regime of bosonic atomic gases in
a quasi-1D geometry.  The physical properties of 1D bosons in this
limit  can be investigated in detail since their 
wavefunction is known explicitly in terms of the one of a
noninteracting fermions in the same external potential
\cite{Girardeau60-65}. In fact, the density profile, the thermodynamic
properties, the collective excitation spectrum and the density
correlation functions coincide with those of an ideal Fermi gas,
leading to interesting manifestations of fermionization of a Bose
gas, such as broadening of the density profiles \cite{Dunjko2001},
an increase of the frequency of collective excitations
\cite{Vignolo_Menotti} and a dramatic reduction of the three-body
recombination rate \cite{Gangardt}.  However the one-body
density matrix, and consequently the momentum distribution, differs
considerably from that of a Fermi gas, due to the phase
correlations stemming  from the bosonic statistics of the Tonks gas.
For the homogeneous case the momentum distribution $n(p)$ at the
origin has a $1/\sqrt{p}$ peak \cite{Momentumdistrold} and for a
harmonically trapped gas the population of the lowest single-particle
state scales as $\sqrt{N}$ with $N$ being the particle number
\cite{Papenbrock2003,Forrester2003}; this shows that due to the strong
interactions the bosons do not form a Bose-Einstein condensate.  In
both cases at large momenta $p>\hbar n$, with $n$ being the density at
the center, the momentum distribution shows characteristic
slowly decaying tails $n(p) \sim p^{-4}$~\cite{Minguzzi_Olshanii}.

Experiments on cold atomic gases under optical confinement in a
quasi-one dimensional geometry are now starting to explore the
strongly interacting regime, demonstrated by the examination of its
correlation properties \cite{Phillips} and of the frequency of
collective modes \cite{Esslinger}. More recently there has been a
significant progress towards the Tonks limit, in which the momentum 
distribution \cite{Paredes2004} and the
thermodynamical properties \cite{Kinoshita2004} have been measured.
Experiments addressing the dynamics of the Tonks gas seem also in
view.  Some aspects of the dynamical evolution of a Tonks gas have
already been theoretically studied. These include the formation of
solitons in a ring geometry \cite{Girardeau2000a}, the splitting and
recombination of a Tonks beam across an obstacle \cite{Girardeau2000b}
and  1D expansion \cite{Ohberg2002}. In the last case, using numerical
calculations on a lattice, the momentum distribution was found to
approach that of an ideal Fermi gas \cite{Rigol2004}.

The aim of this work is to study the dynamical evolution of a
harmonically trapped Tonks gas induced by arbitrary time variations of
the trap frequency. We show that this evolution can be described exactly using 
time-dependent coherent states (see e.g. \cite{Perelomov}), 
in close analogy of the dynamics of Bose-Einstein 
condensates \cite{Kagan1996}. 
These have been investigated also for the
 case of a strongly correlated Bose gas
interacting with an inverse-square pair potential \cite{Sutherland}.
Here we explore in particular how the bosonic or fermionic properties 
of the Tonks gas manifest themselves in  the dynamics of the
coherence.  For this purpose we choose the
momentum distribution as the observable. We first determine explicitly its time
evolution in terms of the initial-time configuration with the aid of a
scaling transformation. We then use this result to study  two
important examples: (i) the 1D expansion of the gas, where we explain
why the gas develops a Fermi shape of the momentum distribution, and
(ii) the large-amplitude breathing modes, where we find the absence of
damping and a rich dynamical evolution in momentum space, displaying
alternating bosonic and fermionic character.

{\em Scaling transformation}.
We consider  $N$ impenetrable bosons with mass $m$  in a
1D geometry at zero temperature and subjected to a harmonic 
potential  
\begin{equation}
\label{potential}
V_{ext}(x,t)=m\omega^2(t) x^2/2 
\end{equation}
with $\omega(t)$ arbitrary
time-dependent trapping frequency and
$\omega(t\le 0)=\omega_0$. We proceed to derive an exact analytic
expression for the  evolution of the Tonks gas wavefunction, the
one-body density matrix and the momentum distribution at all times.  
We start by employing the time-dependent Bose-Fermi mapping
\cite{Girardeau2000a}, 
 which allows to write the many-body wavefunction 
$\Phi_T(x_1,x_2,\ldots,x_N;t)$  
for the Tonks gas in terms of the wavefunction
$\Phi_F(x_1,x_2,\ldots,x_N;t)=(1/\sqrt{N!})\det_{j,k=1}^N \phi_j(x_k,t)$   
of a noninteracting Fermi gas experiencing the same external
potential. The single-particle orbitals $\phi_j(x,t)$ satisfy the
time-dependent Schr{\" o}dinger equation  
\begin{equation}
\label{tdse}
i\hbar \frac{\partial}{\partial t}
\phi_j(x,t)=-\frac{\hbar^2}{2m}\frac{\partial^2}{\partial x^2}\phi_j(x,t)+V_{ext}(x,t)\phi_j(x,t).
\end{equation}
The Tonks gas wavefunction is then constructed by applying to $\Phi_F$
the unit antisymmetric function ${\cal A}
(x_1,\ldots,x_N)=\Pi_{1 \le j< k \le N}{\rm sgn}(x_j-x_k)$,
\begin{equation}
\label{eq:Tonks}
\Phi_T(x_1,\ldots,x_N;t)={\cal A}
(x_1,\ldots,x_N)\Phi_F(x_1,\ldots,x_N;t),
\end{equation}
and is thus properly symmetrized. 

For the case of a time-dependent potential (\ref{potential}) the
introduction of a scaling transformation for both the space and time
coordinates provides an exact solution for
Eq.~(\ref{tdse}), which reads \cite{Perelomov,Kagan1996}
\begin{equation}
\label{eq:scaling}
 \phi_j(x,t) = \frac{1}{\sqrt{b}}\phi_j\left(\frac{x}{b},0\right)
    \exp{\left[i\frac{mx^2}{2\hbar}\frac{\dot{b}}{b}-iE_j\tau
    (t)\right]}.
\end{equation}
In Eq.~(\ref{eq:scaling}) above the scaling factor $b(t)$ obeys the
second-order differential equation
\begin{equation}
  \label{eq:bt}
  \ddot{b}+\omega^2 (t) b =\omega^2_0/b^3
\end{equation}
with initial conditions $b(0)=1$ and $\dot{b}(0)=0$, the rescaled time
parameter is determined by $\tau(t)=\int_0^t dt'/b^2(t')$, and
$\phi_j(x,0)$ are the well-known wavefunctions of the 1D harmonic
oscillator with frequency $\omega_0$ and eigenvalue $E_j$ expressed
in terms of the Hermite polynomials.
We remark that Eq.~(\ref{eq:scaling}) is the unique time-dependent
solution of the linear Schroedinger equation (\ref{tdse}).

Substituting the one-particle states (\ref{eq:scaling}) into
Eq.~(\ref{eq:Tonks}) leads to the final result for the time evolution
of the Tonks gas wavefunction in terms of its initial-time expression
\begin{eqnarray}
  \label{eq:scaling_mb}
  \Phi_T(x_1,..,x_N;t) = b^{-N/2} \Phi_T(x_1/b,..,x_N/b;0)\nonumber \\
  \times \exp\left(\frac{i\dot{b}}{b\omega_0}\sum_j
  \frac{x_j^2}{2l^2_0}\right)\exp\left(-i \sum_j E_j \tau\right),
\end{eqnarray}
with $l_0=\sqrt{\hbar/m \omega_0}$.  Here we used that (i) the scaling
(\ref{eq:scaling}) takes place irrespectively of the one-particle
quantum number, and (ii) the unit antisymmetric operator ${\cal A} $
is invariant under the scaling transformation.

Equation~(\ref{eq:scaling_mb}) allows an immediate  derivation of  an
analytic expression for the one-body density matrix, given by its
first-quantized expression $ g_1(x,y;t) = \!N \!\int \!dx_2.. dx_N
\Phi^*(x,x_2,..,x_N;t)\Phi(y,x_2,..,x_N;t)$, in the scaling form
\begin{equation}
  \label{eq:g1_t}
  g_1(x,y;t) = \frac{1}{b} g_1\left(\frac{x}{b},\frac{y}{b};0\right)
  \exp\left(-\frac{i}{b}\frac{\dot{b}}{\omega_0}
  \;\frac{x^2-y^2}{2l^2_0}\right).
\end{equation}
This result shows first of all that during the generic time evolution
described by Eq.~(\ref{potential}) the absolute value of the one-body
density matrix preserves its power-law behavior at large distances
\cite{Forrester2003,Gangardt2004}.  For the case of an expansion this
was noticed in Ref.~\cite{Rigol2004} from numerical simulations on a
lattice.
Secondly, Eq.~(\ref{eq:g1_t}) yields the exact evolution of the Tonks
gas density profile $\rho(x;t)\equiv g_1(x,x;t)=\rho(x/b;0)/b$, thus
generalizing the result of Ref.\cite{Ohberg2002} for any
time-evolution of the trap frequency.  We remark that Eq.~(\ref{eq:g1_t})
also holds at any finite temperature, as follows from the statistical
Bose-Fermi mapping theorem \cite{GirardeauT}.

The dynamical phase in Eq.~(\ref{eq:g1_t}) enters the expression for
  the time evolution of the momentum distribution $n(p,t) = \int dx dy
  \,e^{ip(x-y)/\hbar} g_1 (x,y;t)$, which upon change of integration
  variables $x/b,\;y/b\to x,y$ reads
\begin{eqnarray}
  \label{eq:momentum_t}
  n(p,t) &=& b\int \!dx dy\;g_1(x,y;0) \nonumber \\&\times&
  \exp\left[-ib\left(\frac{\dot{b}}{\omega_0}\frac{x^2-y^2}{2l_0^2}
  -\frac{p(x-y)}{\hbar}\right)\right].
\end{eqnarray}
Below we describe two cases where the dynamical phase acquired by the
one-body density matrix strongly influences the time behavior of the
momentum distribution.

\begin{figure}
  \centering
 \psfrag{t}{$t$} \psfrag{p}{$p l_0/\hbar$}
\psfrag{n(p)}{$n(p,t)/l_0 $}
\includegraphics[width=7.5cm,angle=270]{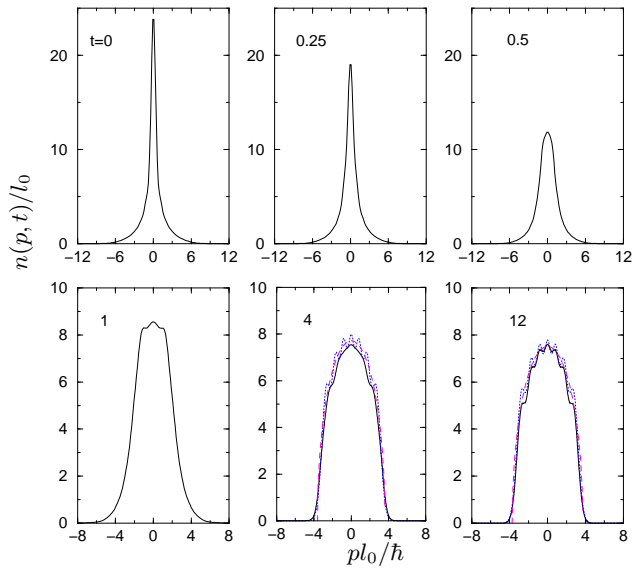} \vspace{0.2cm}
  \caption{Momentum distribution of an expanding Tonks gas with $N=7$ at different times (in units of $1/\omega_0$) as indicated on the panels, from numerical solution (solid lines), asymptotic fermionic limit (dotted lines) and Thomas-Fermi approximation (dashed lines). The units are indicated on the axis labels.}
  \label{fig:expansion}
\end{figure}

{\em Expansion.}  A 1D expansion could be achieved in an experiment by
turning off only the longitudinal confinement.  To describe this case
we set $\omega(t\le 0)=\omega_0$ and $\omega(t)=0$ for $t>0$, and the
solution of Eq.~(\ref{eq:bt}) for the scaling parameter $b(t)$ is
given by $b(t)=\sqrt{1+\omega_0^2t^2}$. As this parameter becomes
increasingly large with time, we are able to determine analytically
the behavior of the momentum distribution (\ref{eq:momentum_t}) for
long times by the method of stationary phase. The points for which the
phase is stationary are given by $x^*=y^*=\omega_0 p l_0^2/(\dot b
\hbar)$, and hence we find that the momentum distribution is
asymptotically determined solely by the diagonal part of the
equilibrium one-body density matrix, i.e., the particle density
profile $\rho(x;0)$ which is identical to the one of an ideal Fermi
gas.
For the case of harmonic confinement the latter is proportional to the
equilibrium momentum distribution $n_F(p)$ of the ideal Fermi gas
\cite{note1} thus allowing to write the final expression for the Tonks
gas momentum distribution as
\begin{equation}
\label{eq:fermi_edge}
n(p,t)= |\omega_0/\dot{b}| \,\, n_F(\omega_0p/\dot{b}).
\end{equation}
This result can be also understood as follows. Since the initial
momentum distribution of the atoms is very narrow, the final momentum
distribution is determined by the hydrodynamic velocity field acquired
during expansion. Such a velocity field is obtained from the dynamical
phase in the Eq.(\ref{eq:g1_t}) and is linear in the position. Hence 
its distribution has the same shape as the particle density profile of 
a Fermi gas.
For $N\gg 1$ Eq.~(\ref{eq:fermi_edge}) is well described by the
Thomas-Fermi approximation, which neglects the quantum shell
oscillations of order $1/N$ \cite{mehta},
\begin{equation}
  \label{eq:momentTF}
n(p,t) \simeq 2(l_0^2/\hbar) |\omega_0/\dot b| \,
\sqrt{P_F^2-(\omega_0 p/\dot b)^2}.
\end{equation}
with $P_F=\hbar \sqrt{2N}/l_0$ being the Fermi momentum.

We estimate the characteristic time, $t_{F}$, for such ``dynamical
fermionization'' by considering that in the Thomas-Fermi regime the
one-body density matrix depends on coordinates only through the
dimensionless ratios $x/R$ and $y/R$. By rewriting
Eq.~(\ref{eq:momentum_t}) in such rescaled coordinates and in terms of
$p/P_F$ we see that the large parameter governing the dynamic phase
in Eq.~(\ref{eq:momentum_t}) is $N b \dot b /\omega_0$, and thus we
find $t_{F}\sim 1/N\omega_0\sim \hbar/E_F$, $E_F$ being the Fermi
energy.

Equation (\ref{eq:fermi_edge}) provides an accurate description of the
momentum distribution for long times and for small momenta $p\le
P_F$. We proceed now to derive the second exact result regarding the
large momentum behavior of $n(p,t)$ at all times. From the scaling
solution (\ref{eq:scaling_mb}) it follows that during its time
evolution the many-body wavefunction displays the same type of cusp
singularity as in its equilibrium configuration. Since no additional
singularity is present in the dynamical phase, this cusp determines
alone the large momentum behavior of the momentum distribution, as in
the case of the equilibrium solution \cite{Minguzzi_Olshanii}. Hence,
we find a power-law decay of the momentum distribution at large $p$,
with an additional time-dependent suppression factor which originates
from the dilatation of the interparticle distances during the
expansion,
\begin{equation}
n(p,t)\sim p^{-4} b^{-3}.
\label{eq:tails} 
\end{equation}
This complements the result (\ref{eq:fermi_edge}) showing that at time
$t\ge 1/\omega_0$ the fermionization of the momentum distribution is
complete and the large-$p$ tails are negligible.

We have tested the above predictions and explored the expansion at
early times by numerically evaluating the momentum distribution
Eq.~(\ref{eq:momentum_t}) with a Fast Fourier Transform method using
the explicit expression of the initial time one-body density matrix in
terms of a determinant of Hankel type~\cite{Forrester2003}.
The results, reported in Fig.~\ref{fig:expansion}, show that a broad
Fermi distribution rapidly develops during expansion, followed by
small adjustments of the quantum shell oscillations, while the
$p^{-4}$ tails become less and less important at long expansion times.

\begin{figure}
  \centering \psfrag{n(p)}{$n(p,t)/l_0$}
 \psfrag{p/pho}{$p l_0/\hbar$} \psfrag{t}{$t$}
\includegraphics[width=8.2cm,angle=0]{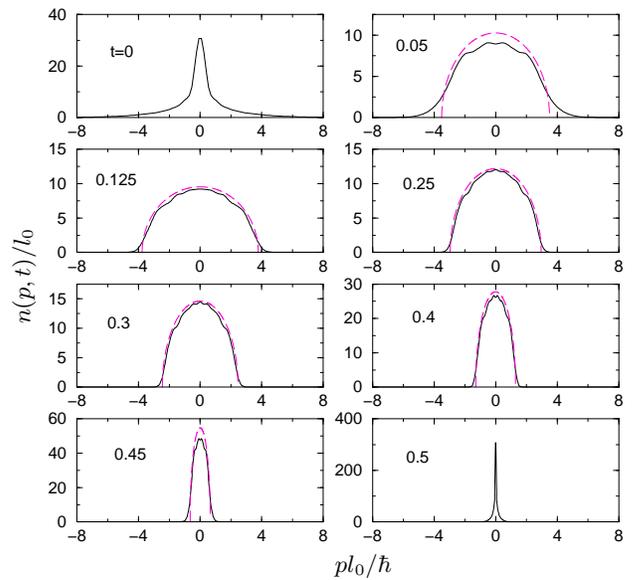}
  \caption{Momentum distribution of an oscillating  Tonks gas with $N=9$ and  $\omega_0/\omega_1=10$ at different times (in units of $T=\pi/\omega_1$) indicated on the panels,  from numerical solution (solid lines) and Thomas-Fermi approximation (dashed lines). The units are indicated on the axis labels.} 
  \label{fig:oscillation}
\end{figure}

{\em Oscillations.} As a second example, we consider now the case of
an abrupt change of the trap frequency which induces large-amplitude
``breathing'' oscillations in the gas, i.e.~we set $\omega(t\le
0)=\omega_0$ and $\omega(t)=\omega_1$ for $t>0$, with
$\omega_1<\omega_0$. The solution of Eq.~(\ref{eq:bt}) is
\begin{equation}
  \label{eq:btomega01}
  b(t) = \sqrt{1 +(\omega_0^2-\omega_1^2)\sin^2(\omega_1
  t)/\omega_1^2}
\end{equation}
which describes periodic oscillations between $1$ and
$\omega_0/\omega_1$ with period $T=\pi/\omega_1$.

The dynamical evolution of the cloud in coordinate space is described,
according to Eq.~(\ref{eq:g1_t}), by the self-similar 
breathing of the density profile with a time law given in
Eq.~(\ref{eq:btomega01}). Notice that the solution
(\ref{eq:btomega01}) implies that the oscillation is undamped. 
The time evolution in momentum space, given
by Eq.~(\ref{eq:momentum_t}), displays a richer structure, and in
particular an oscillating behavior between a bosonic-like and
fermionic-like momentum distribution which is illustrated in
Fig.~\ref{fig:oscillation}.  The main features of the dynamics may be
understood by the following analytical considerations.  When the
condition $N b \dot b /\omega_0>1$ holds, the stationary phase method
can be employed, yielding that a time-dependent Fermi-like structure
develops as in Eq.~(\ref{eq:fermi_edge}), and that the
large-wavevector tails are suppressed as in Eq.~(\ref{eq:tails}). In
the regime $N\gg1 $ and $\omega_0 \gg \omega_1$ this result is valid
for most values of $t$, and closely resembles  the dynamics of an ideal Fermi
gas. The latter is given by the exact expression \cite{notaFermi}
\begin{equation}
 \label{eq:momentum_free_fermions_res}
  n(p,t) = B(t) n_F(B(t)p) .
\end{equation}
with $B=b/\sqrt{1+b^2\dot{b}^2/\omega_0^2}$ being related to the
scaling of the kinetic energy ($B(t)=1$ for the expansion).
In a small time interval $\omega_0 \Delta
t=N^{-1}\omega_0^2/(\omega_0^2-\omega_1^2)$ around the turning point
$t=T/2$ the fermionic description does not hold. There, using
Eq.~(\ref{eq:momentum_t}) we find that the atomic cloud recovers the
initial bosonic momentum distribution rescaled by a factor
$b_{max}=\omega_0/\omega_1$, according to $n(p,T/2)=b_{max}
n(b_{max}p,0)$.

In conclusion,
we have shown that the dynamical phase acquired by the many-body
wavefunction during the evolution is responsible for the ``dynamical
fermionization'' found in the 1D expansion and that during a
large-amplitude breathing mode the cloud displays oscillations in
momentum space between a Fermi-like and a Bose-like structure.  

In this work we have considered only the Tonks-Girardeau limit of the
Lieb-Liniger model. In this case the scaling transformation is exact
and predicts a coherent, undamped motion of the cloud, which is a very
peculiar feature of the Tonks regime. We suggest that the absence of
damping in the breathing modes may be used to characterize the truly
Tonks regime: for finite values of the coupling strength we expect
instead a damped motion of the oscillations.  Our predictions directly
apply to the experiments on ultracold atomic gases in tight optical
traps, where the time evolution of the momentum distribution should be
experimentally accessible by allowing a 3D expansion of the cloud
\cite{Paredes2004}.  The possibility of manipulating coherently the
Tonks gas and the absence of damping may make this system interesting
for the applications to quantum information.


{\em Acknowledgments} We thank E. Bogomolny, O. Bohigas, S. Brazovsky,
L.P. Pitaevskii and G.V. Shlyapnikov for valuable comments and
discussions.  We acknowledge support from the Centre National de la
Recherche Scientifique (CNRS) and from the Minist\`ere de la Recherche
(grant ACI Nanoscience 201). A.M. acknowledges financial support from
SNS Pisa.  LPTMS is a mixed research unit (UMR 8626) of CNRS and
Universit\'e Paris Sud.




\begin{thebibliography}{99}
\bibitem{note1} During the expansion the momentum distribution of an
ideal Fermi gas is stationary, see the discussion after
Eq.~(\ref{eq:momentum_free_fermions_res}).
\end{thebibliography}
\end{document}